\documentclass[12pt,epsf,epsfig]{article}
\usepackage{amsmath}

\begin{document}

\begin{center}
{\huge{Relieving String Tension By Making Baby
Universes in a Dynamical String Tension Braneworld Model }} 
 \footnote{Honorable mention in the Gravity Research Foundation 2022  Essays on Gravitation Competition, submission date : March 29th, 2022,  e-mails: E.G., corresponding author, guendel@bgu.ac.il, and Z.M., merali@fqxi.org }  \\
\end{center}

\begin{center}
 Eduardo I. Guendelman  \\
\end{center}

\begin{center}
\ Department of Physics, Ben-Gurion University of the Negev, Beer-Sheva, Israel \\
\end{center}

\begin{center}
\ Frankfurt Institute for Advanced Studies, Giersch Science Center, Campus Riedberg, Frankfurt am Main, Germany \\
\end{center}

\begin{center}
\ Bahamas Advanced Studies Institute and Conferences,  4A Ocean Heights, Hill View Circle, Stella Maris, Long Island, The Bahamas \\
\end{center}
        
\begin{center}
 Zeeya Merali  \\
\end{center}
\begin{center}
\ Foundational Questions Institute (FQXi), PO Box 3055, Decatur, GA 30031,
USA \\
\end{center}
\abstract
String tension fundamentally determines the properties of strings; yet its value is
often assigned arbitrarily, creating a fine-tuning problem. We describe a mechanism
for dynamically generating string tension in a flat or almost flat spacetime, using the
‘modified measures formalism,’ which in turn naturally generates a new type of
stringy brane-world scenario. Such a scenario allows strings to achieve near infinite
tension confining the strings to two very close expanding surfaces, but the infinite tensions also threatens to distort the near-flat embedding spacetime through large back
reactions. We argue that this danger can be neutralised via the creation of a baby
universe—a growing region of emdedding spacetime that divorces from the ambient embedding
spacetime, while our universe is still a brane separating two nearly flat spacetimes. The avoidance of a minimum length and a maximum Hagedorn
temperature in the context of dynamical string tension generation are also discussed.

\newpage

\section{Introduction}

The tension of the string is a fundamental parameter in string theory, playing a
crucial role in determining the vibration modes of the string \cite{stringtheory}; yet in many models
its value is arbitrary, creating a fine-tuning problem. In previous publications \cite{a,c,supermod, cnish, T1, T2, T3}, see also the related treatment by Townsend and collaborators, \cite{xx,xxx}. 
we have demonstrated that, rather than putting it in by hand, the string tension can
be dynamically generated.  We  used the modified measures formalism, which was  previously used for a certain class of modified gravity theories under the names of Two Measures Theories or Non Riemannian Measures Theories, see for example \cite{d,b, Hehl, GKatz, DE, MODDM, Cordero, Hidden}
Leads to the  modified measure approach of conventional string theory \cite{stringtheory}, where  rather than to put the string tension by hand it appears dynamically.
In \cite{Ansoldi} and \cite{ESSAY} 
we have also introduced the ¨tension scalar¨, which is an additional
background
field that can be introduced into the theory for the bosonic case (and expected to be well defined for all types of superstrings as well) that changes the value of the tension of the extended object along its world sheet. 
This approach was then applied to the construction of braneworlds  
in  \cite{braneworldswithDT}, and before studying issues that are very special of this paper we review some of the material contained in previous papers,  first present the string theory with a modified measure and containing also gauge fields that like in the world sheet, the integration of the equation of motion of these gauge fields gives rise to a dynamically generated string tension, this string tension may differ from one string to the other.

In exploring the implications of a dynamical string tension for braneworlds in \cite{braneworldswithDT} a
number of unexpected and potentially interesting factors became apparent: First,
when considering the simplest non-trivial case of two types of strings, it was found
that the mechanism naturally generates a new type of stringy brane world scenario.
Second, upon investigation into whether a string with near infinite tension might
cause large back reactions that distort flat spacetime, it was discovered that this
problem could be alleviated by invoking a mechanism developed in a seemingly
unrelated context, namely, the creation of baby universes in an inflationary scenario \cite{BlauGuendelmanGuth}, \cite{FromBlackHolestoBabyUniverses}
. This raises the question of whether universe creation from flat, or almost flat,
space is a necessary consequence of a model with a dynamically-generated string
tension braneworld models. Third, we argue that in dynamical tension string theories, at locations
where the string tensions diverge, there will be no minimum length or maximum
Hagedorn temperature. Here we review these findings and examine their
implications.

\section{Tension Scalar and String Dependent Metrics }

As we discussed in \cite{braneworldswithDT}, we can incorporate the result of the tension as a function of scalar $\phi$, the tension field, given as $e\phi+T_i$, for a string with the constant of integration $T_i$ by defining the action that produces the correct 
equations of motion for such string, adding also other background fields, the anti symmetric  two index field $A_{\mu \nu}$ that couples to $\epsilon^{ab}\partial_a X^{\mu} \partial_b X^{\nu}$
and the dilaton field $\varphi $ that couples to the topological density $\sqrt{-\gamma} R$
\begin{equation}\label{variablestringtensioneffectiveacton}
S_{i} = -\int d^2 \sigma (e\phi+T_i)\frac12 \sqrt{-\gamma} \gamma^{ab} \partial_a X^{\mu} \partial_b X^{\nu} g_{\mu \nu} + \int d^2 \sigma A_{\mu \nu}\epsilon^{ab}\partial_a X^{\mu} \partial_b X^{\nu}+\int d^2 \sigma \sqrt{-\gamma}\varphi R .
\end{equation}
Notice that if we had just one string, or if all strings will have the same constant of integration $T_i = T_0$.

In any case, it is not our purpose here to do a full generic analysis of all possible background metrics, antisymmetric two index tensor field and dilaton fields, instead, we will take  cases where the dilaton field is a constant or zero, and the antisymmetric two index tensor field is pure gauge or zero, then the demand of conformal invariance for $D=26$ becomes the demand that all the metrics
\begin{equation}\label{tensiondependentmetrics}
g^i_{\mu \nu} =  (e\phi+T_i)g_{\mu \nu}
\end{equation}
will satisfy simultaneously the vacuum Einstein´s equations,
Notice that if we had just one string, or if all strings will have the same constant of integration $T_i = T_0$, then all the 
$g^i_{\mu \nu}$ metrics are the same and then 
(\ref{tensiondependentmetrics}) is just a single field redefinition and therefore there will be only one metric that will have to satisfy Einstein´s equations, which of course will not impose a constraint on the tension field $\phi$ . 
\section{Determination of the Tension Scalar for a
System of Strings with Multiple Tensions }
The interesting case to consider is therefore many strings with different $T_i$. Each string is then considered as an independent system that can be quantized.
We take into account the string generation by introducing the tension as a function
of the scalar field, as a factor inside a Polyakov type action with such string
tension. Let us consider the simplest case of two types of strings, labeled $1$ and $2$ with  $T_1 \neq  T_2$ , then we will have two Einstein´s equations, for $g^1_{\mu \nu} =  (e\phi+T_1)g_{\mu \nu}$ and for  $g^2_{\mu \nu} =  (e\phi+T_2)g_{\mu \nu}$, 

\begin{equation}\label{Einstein1}
R_{\mu \nu} (g^1_{\alpha \beta}) =  R_{\mu \nu} (g^2_{\alpha \beta}) = 0
\end{equation}

These two simultaneous conditions above  impose a constraint on the tension field
 $\phi$, because the metrics $g^1_{\alpha \beta}$ and $g^2_{\alpha \beta}$ are conformally related, but Einstein's equations are not conformally invariant, so the condition that  Einstein's equations hold  for both  $g^1_{\alpha \beta}$ and $g^2_{\alpha \beta}$
is highly non trivial. Then for these situations, we have,
\begin{equation}\label{relationbetweentensions}
e\phi+T_1 = \Omega^2(e\phi+T_2)
\end{equation}
 which leads to a solution for $e\phi$
 
\begin{equation}\label{solutionforphi}
e\phi  = \frac{\Omega^2T_2 -T_1}{1 - \Omega^2} 
\end{equation}
which in turn leads to the tensions of the different strings to be
\begin{equation}\label{stringtensions}
 e\phi+T_1 = \frac{\Omega^2(T_2 -T_1)}{1 - \Omega^2} , \\\\\
 e\phi+T_2 = \frac{(T_2 -T_1)}{1 - \Omega^2} 
\end{equation}
Both tensions are positive if $T_2 -T_1$ is positive and $\Omega^2$ is also positive and less than $1$.

\subsubsection{Flat space in Minkowski coordinates and flat space after a special conformal transformation }

Let us s now study now a case where the two spaces are related by a conformal transformation, which will be flat space in Minkowski coordinates and flat space after a special conformal transformation.The flat space in Minkowski coordinates is,

 \begin{equation}\label{Minkowskii}
 ds_1^2 = \eta_{\alpha \beta} dx^{\alpha} dx^{\beta}
\end{equation}
 $ \eta_{\alpha \beta}$ is the standard Minkowskii metric, with 
$ \eta_{00}= 1$, $ \eta_{0i}= 0 $ and $ \eta_{ij}= - \delta_{ij}$.
This is of course a solution of the vacuum Einstein´s equations.
We now consider the conformally transformed metric

 \begin{equation}\label{Conformally transformed Minkowski}
 ds_2^2 = \Omega(x)^2  \eta_{\alpha \beta} dx^{\alpha} dx^{\beta}
\end{equation}
where we will choose this conformal factor to be that obtained from a special conformal transformation, see \cite{braneworldswithDT} for the $D$ dimensional case and \cite{Culetu} for the  $D=4$ case.
\begin{equation}\label{ special conformal transformation}
x ^{\mu}\prime =  \frac{(x ^{\mu} +a ^{\mu} x^2)}{(1 +2 a_{\nu}x^{\nu} +   a^2 x^2)}
 \end{equation}
in summary, we have two solutions for the Einstein´s equations,
 $g^1_{\alpha \beta}=\eta_{\alpha \beta}$ and 
 
 \begin{equation}\label{ conformally transformed metric}
 g^2_{\alpha \beta}= \Omega^2\eta_{\alpha \beta} =\frac{1}{( 1 +2 a_{\mu}x^{\mu} +   a^2 x^2)^2} \eta_{\alpha \beta}
 \end{equation}
 
 We can then study the evolution of the tensions using 
 $\Omega^2 =\frac{1}{( 1 +2 a_{\mu}x^{\mu} +  a^2 x^2)^2}$.
 We will consider only $a^2 \neq 0 $ for the purpose of this essay (for $a^2 =0 $ the brane world consists of two parallel surfaces moving with the speed of light, and the strings being confined between these two surfaces since at the surdaces themselves the string tensions become infinite  \cite{braneworldswithDT} ).
 
  \section{ Brane Worlds  in Dynamical  String  Tension Theories }
  We now consider the case when $a^\mu$ is not light like, so that $a^2 \neq 0$. Regardless of whether  $a^\mu$ is space like or time like, we will have thick  Brane Worlds  where strings can be constrained  between two concentric spherically symmetric bouncing higher dimensional spheres and where the distance between these two  concentric spherically symmetric bouncing higher dimensional spheres approaches zero at large times.
  The string tensions of the strings one and two are given by
    \begin{equation}\label{stringtension1forBraneworld}
 e\phi+T_1 = \frac{(T_2-T_1)( 1 +2 a_{\mu}x^{\mu} +  a^2 x^2)^2}{( 1 +2 a_{\mu}x^{\mu} +  a^2 x^2)^2-1}=  \frac{(T_2-T_1)( 1 +2 a_{\mu}x^{\mu} +  a^2 x^2)^2}{(2 a_{\mu}x^{\mu} +  a^2 x^2)(2+2 a_{\mu}x^{\mu} +  a^2 x^2)}
\end{equation}
  \begin{equation}\label{stringtension2forBraneworld}
 e\phi+T_2 = \frac{(T_2-T_1)}{( 1 +2 a_{\mu}x^{\mu} +  a^2 x^2)^2-1}=  \frac{(T_2-T_1)}{(2 a_{\mu}x^{\mu} +  a^2 x^2)(2+2 a_{\mu}x^{\mu} +  a^2 x^2)}
\end{equation}
Then, the locations where the string tensions go to infinity are given by,

\begin{equation}\label{boundariesforBraneworld1}
2 a_{\mu}x^{\mu} +  a^2 x^2 = 0
\end{equation}
or 
\begin{equation}\label{boundariesforBraneworld2}
2 +2 a_{\mu}x^{\mu} +  a^2 x^2 = 0
\end{equation}
Let us start by considering the case where  $a^\mu$ is time like, then without loosing generality we can take  $a^\mu = (A, 0, 0,...,0)$. Then the denominator in (\ref{stringtension1forBraneworld}) , (\ref{stringtension2forBraneworld}) is
\begin{equation}\label{denominatortimelike}
(2 a_{\mu}x^{\mu} +  a^2 x^2)(2+2 a_{\mu}x^{\mu} +  a^2 x^2) =
(2At +A^2(t^2-x^2))(2+2At++A^2(t^2-x^2))
\end{equation}

The condition (\ref{boundariesforBraneworld1}) implies then that
\begin{equation}\label{bubbleboundaryforBraneworld1a}
 x^2_1  + x^2_2 + x^2_3.....+ x^2_{D-1}- (t+ \frac{1}{A})^2 = -\frac{1}{A^2}
\end{equation}
while the other boundary of infinite string tension (\ref{boundariesforBraneworld2}) is given by,
\begin{equation}\label{bubbleboundaryforBraneworld1b}
 x^2_1  + x^2_2 + x^2_3.....+ x^2_{D-1}- (t+ \frac{1}{A})^2 = \frac{1}{A^2}
\end{equation}
So we see that (\ref{bubbleboundaryforBraneworld1b}) represents an exterior boundary which has an bouncing  motion with a minimum radius $\frac{1}{A}$ at $t = - \frac{1}{A}$ , 
The denominator (\ref{denominatortimelike}) is positive between these two bubbles. This bounce can be the starting point for a Universe creation process.
For $T_2 -T_1$ positive the tensions are positive and diverge at the surfaces (\ref{bubbleboundaryforBraneworld1a})  and (\ref{bubbleboundaryforBraneworld1b}).
The internal boundary (\ref{bubbleboundaryforBraneworld1a}) exists only for times $t$ smaller than $-\frac{2}{A}$ and bigger than  
$0$, so in the time interval $(-\frac{2}{A},0)$
there is no inner surface of infinite tension strings.
This inner surface collapses to zero radius at  $t=-\frac{2}{A}$
and emerges again from zero radius at $t=0$. 
For large positive or negative times, the difference between the upper radius  and the lower radius goes to zero as  $t \rightarrow \infty$

\begin{equation}\label{asymptotic}
\sqrt{\frac{1}{A^2} +(t+ \frac{1}{A})^2 } -\sqrt{-\frac{1}{A^2} +(t+ \frac{1}{A})^2 }\rightarrow \frac{1}{t A^2}\rightarrow 0  
\end{equation}
of course the same holds  $t \rightarrow -\infty$.
This means that for very large early or late times the segment where the strings would be confined (since they will avoid having infinite tension) will be very narrow and the resulting scenario will be that od a brane world for late or early times, while in the bouncing region the inner surface does not exist.

The case when   $a^\mu$ is space like leads to a similar picture, so it will not be discussed separately, details of this case are worked out in \cite{braneworldswithDT} 
   \section{ Baby Universe Creation As a Mechanism
To Resolve the Backreaction of Infinite Tension on the Flat Space
Backgrounds}

Now the interesting question appears: 
Our whole construction of the braneworld has been based on the conformal mapping between two flat spaces and in principle it represents a vacuum solution where test strings acquire string tensions that diverge at two concentric and expanding surfaces.
Furthermore, as we start to populate the braneworld with actual strings, these strings will have infinite tension at the borders of the braneworld. A natural question one may ask at this point is the following :  Are the flat space backgrounds of our construction consistent with the presence of very high Tension Strings or will the backreaction from the very large string tension destroy this basic feature of the model ?.  This question requires a non trivial investigation because the presence of arbitrarily large string tensions would appear to generate  substantial back reaction from the space time and possibly large deviations from the construction based on the flat spaces in the previous sections, but is that so? . As shown in \cite{GuendelmanPortnoy2022},  it appears that the introduction of large tension strings is consistent with a general relativistic picture of Universe Creation from flat or Almost Flat space , where two flat or almost flat spaces can be matched through a membrane with the matter content of a gas of strings with arbitrarily large string tension. This is the scenario studied by us  in 1+1 membranes in a 2+1 embedding space \cite{Jacob1} and in for 2+1 membranes in a 3+1 embedding space \cite{universesfromflatspace}  and has been now generalized for a 1+(D-2) membrane moving in a  1+(D-1) space  \cite{GuendelmanPortnoy2022} , the membrane expands to infinty as the tensions are infinite. 

   \subsection{A General Relativistic ¨Macroscopic¨ String Gas Shells Model With Arbitrarily Large Tensions }
   Now we will see in more details how two spaces that are almost flat can be matched with a surface with matter 
described by a string gas with arbitrarily large tensions \cite{universesfromflatspace}. There are obstacles to directly compare the braneworld solutions in dynamical tension strings and those found in \cite{universesfromflatspace}, which are: 1) \cite{universesfromflatspace} describes a $2+1$ dimensional brane moving in an embedding bulk space of $3+1$ dimensions, while for string theories we must consider higher dimensions, 2) in \cite{universesfromflatspace} Einstein gravity is assumed with one metric to hold in the embedding bulk space, while the effective gravity theory for the dynamical tension strings two string metrics appear, 3) in \cite{universesfromflatspace} an infinitely thin brane is considered, while in  dynamical tension strings the branes are thick, that is why the thin wall model will be referred as a ¨macroscopic¨ representation of the braneworld scenario.
A difference can be resolved in a simple way is generalizing the dimensions of the brane to  $(D-2)+1$ and that of the embedding space to $(D-1)+1$. This we will do, while we hope the other aspects will not change the basic qualitative aspects of the comparison.

We consider then a surface or thin shell with  $D-2$ spacial dimensions, where in this shell a gas of strings with the equation of state that relates the surface pressure $p$ to the  $\sigma$ being
\begin{equation}\label{string gas}
p= - \frac{\sigma}{D-2}
\end{equation}
see for example a discussion of the string gas equation o state in $4D$ cosmology in \cite{stringgasequationofstate})
and for an example involving string gas shells see \cite{StringGasShells},
so for $D=3$, we obtain that the surface becomes a line with $p= - \sigma $, This was a matching corresponding to a particular choice of the ones studied in 
\cite{Jacob1}, while the  $D=4$ corresponds to a membrane ($2+1$ dimensional brane) moving in $3+1$ universe with a string gas matter in it \cite{universesfromflatspace}.

applying a local conservation law of the energy momentum in the brane defined by eq. (\ref{string gas}) leads to the possibility of integrating $\sigma$,
\begin{equation}\label{sigma}
\sigma= \frac{\sigma_0}{r^{D-3}}
\end{equation}
where $\sigma_0$ is a constant. As we can see, for $D=3$, $\sigma=$ constant, as  considered as a particular case in \cite{Jacob1} while for $D=3$, $\sigma= \frac{\sigma_0}{r}$ as  considered  in \cite{universesfromflatspace}.
These cases used the Israel matching conditions \cite{Israel} for two space times separated by the string gas shell that we will now generalize to higher dimensions. Following \cite{universesfromflatspace} generalized now to higher dimensions, we consider two stationary metrics with rotational invariance of the form,
\begin{equation}\label{metrics}
ds_{+}^2= -A_{+}dt^2 +\frac{dr^2}{A_{+}} +r^2 d\Omega^2_{D-2},\\\ 
ds_{-}^2= -A_{-}dt^2 +\frac{dr^2}{A_{-}} +r^2 d\Omega^2_{D-2}
\end{equation}
for the inside and outside metrics. Here $d\Omega^2_{D-2}$ represents the contribution to the metric of the $D-2$ angles relevant to the spherically symmetric solutions in $D$ space time dimensions.
$A_{+}$ and $A_{-}$ are functions of $r$ , different for the inside and the outside, matched at a bubble defined by a trajectory
$r= r(\tau)$, 
then the matching condition as a consequence of the Israel analysis \cite{Israel} generalized to $D$ dimensions reads,
\begin{equation}\label{matching}
\sqrt{A_{-} + \dot{r}^2}-\sqrt{A_{+} + \dot{r}^2} = \kappa \sigma r
\end{equation}
where $\kappa$ is proportional to Newton constant in $D$ dimensions.

\subsection{Normal Matching, Wormhole or Baby Universe  Matching and the Braneworld Scenarios }
	We may call normal matching, the situation when both square roots in equation (\ref{matching}) are positive, it corresponds to the situation when the radius $r$ increases as we go from the inside towards the brane, then as we cross the brane, the radius continues to increase, it is easy to see that with normal matching the brane cannot expand to infinity, specially for a gas of very heavy strings.
	
The square roots are not necessarily positive however, the signs can be negative for example for a wormhole matching as has been discussed in details in $D=4$, which corresponds to a membrane ($2+1$ dimensional brane) moving in $3+1$ universe with a string gas matter in it \cite{universesfromflatspace} and corresponds indeed to a baby universe creation. Another case where a difference a sum of the  two terms is obtained, or what is equivalent, we can say that  the second square root is negative is when considering a brane word scenario where the radius growths as we go out from the brane on both sides, see for example  \cite{DynamicsofAnti-deSitterDomainWalls} .
The assignment of signs of the square roots when one of the spaces in a Schwarzschild space can be worked out rigorously by studying the problem using Kruskal–Szekeres  coordinates \cite{BlauGuendelmanGuth}
where these expressions were used for the study of the dynamics of false vacuum bubbles and baby universe creation.

We will now study the case where inside we have flat space, that is
 $A_{-} =1$ and outside a $D$ dimensional Schwarzschild solution with maximal rotational invariance, which gives the Tangherlini solution \cite{Tangherlini} $A_{+} = 1- \frac{c_{1}}{r^{D-3}} $
where $c_{1}$ is a constant. In the Tangherlini solution the radial fall off $\frac{1}{r}$ of the Newtonian potential is replaced by the $\frac{1}{r^{D-3}}$ behavior. These expressions have been used for $D=3$ in \cite{Jacob1}, while the  $D=4$ was studied in  \cite{universesfromflatspace}, we now generalize for any dimension.
\subsection{Effective Potential for the Brane Evolution}
Solving from \ref{matching} for one of the square roots and then solving for the other square root and squaring again, we obtain the particle in a potential like equation,
\begin{equation}\label{particlelikeequation}
\dot{r}^2 + V_{eff}(r)= 0
\end{equation}
where 

\begin{equation}\label{effectivepotentialalternative}
 V_{eff}(r)=1 - (
 \frac{c_1}{2\kappa \sigma_0 r} +  \frac{\kappa \sigma_0}{2r^{D-4}}  )^2
\end{equation}
This expression coincides with the expression obtained in \cite{universesfromflatspace} for the case where $D=4$, as we see from the above expression, the potential, even for $c_1 \neq 0$ goes to a constant
as $r \rightarrow \infty $. This constant can be positive or negative depending on whether 
$\kappa \sigma_0$ is big or small, for $\kappa \sigma_0$  the asymptotic value of the potential is negative and the membrane approaches $\infty $ with constant velocity
regardless of the value of $c_1$, if  $\kappa \sigma_0 $ is bigger than $2$, so large string tensions produce indeed child universes even in the case the inside and outside spacetimes are flat ( the inside space time is always flat, while the outside space is flat for $c_1 = 0$). The matching requires a wormhole as shown in \cite{universesfromflatspace}.

For D bigger than 4 there is a maximum radius given by 
\begin{equation}\label{effectivepotentialalternative}
 V_{eff}(r)= 0
\end{equation}
which, if we take the limit that both internal and external space are flat ($c_1 =0$), this leads to the solution for the maximum radius For $D$ bigger than  $4$,

 \begin{equation}\label{solforraduis}
  r_m= 
 (\frac{(\kappa \sigma_0)^2}{4})^{^{\frac{1}{2D-8}}}= (\frac{\kappa \sigma_0}{2})^{^{\frac{1}{D-4}}}
 \end{equation}
 
 Notice that as $\sigma_0 \rightarrow \infty$ also $r_m\rightarrow \infty$ demonstrating that the shell with superheavy strings does indeed expand to infinity.

For $D=3$ we have that for $ r  \rightarrow 0$ and  $ r  \rightarrow  \infty $
the potential is negative, it is negative everywhere if $c_1 $ is bigger than one, or in the case of more interest to us, if  $c_1 =0$ in the limit $\sigma_0 \rightarrow  \infty$, when it will be negative everywhere except for an infinitesimal region arround $r=0$.
For $D=4$, there is a finite critical value of the string tension where above that the the potential is negative every where regardless of the value of $c_1 $ so the membranes go to infinity \cite{universesfromflatspace}. For $D$ bigger than $4$ the potential is negative as $r \rightarrow 0 $ and approaches the point of maximum expansion $r_m $ (\ref{solforraduis}), which goes to infinity as the tension goes to infinity. 
The detailed study of the matching reveals in all dimensions for high  tensions of the strings reveals a wormhole and an associated baby universe forming in the embedding space , the physical universe is the membrane that separates the two flat or almost flat spaces. The result that a very heavy string gas shell is consistently matching two flat or very nearly flat spaces can be interpreted as a consequence of the gravitational energy of the wormhole space time being negative. 
   \section{Possibility of No Minimum Length and No Hagedorn Temperature  }
String theories have a minimum length  inversely proportional to the string tension \cite{stringuncertaintyprinciple} and a maximum temperature, the Hagedorn temperature, which is  proportional to the string tension. In dynamical tension string theories , at the locations where the string tensions diverge there will be therefore no minimum length or maximum Hagedorn temperature \cite{ESSAY}.

A minimum length and a maximum temperature are indeed two aspects of the same thing, since temperature in the euclidean formulation of finite temperature field theories, is the inverse of the radius of the periodic compactification in the temporal dimension, but if there is a limit as to how small this compactified temporal dimension is, since now time is treated as a spacial direction, and there is a minimum length,  then there must be a maximum temperature. The edges of the braneworld in particular, where the string tensions diverge can therefore have well defined positions and where infinite temperatures can take place as well.

\textbf{Acknowledgments}
 	We thank  Stefano Ansoldi, Steve Blau, Aharon Davidson, Alan Guth,  David Owen, Nobuyuki Sakai, Idan Shilon and Jacob Portnoy for collaborations on the subject of Universe creation, David Benisty, Emil Nissimov, Svetlana Pacheva, Alex Kaganovich, Euro Spallucci and Stefano Ansoldi for collaboration on the modified measure gravity and  string theories,  the Foundational Questions Institute (FQXi) for specially  funding this collaboration through a generous grant. We also thank support from the Quantum Gravity Phenomenology Cost Action,  CA18108.

\end{document}